# InterSliceBoost: Identifying Tissue Layers in Three-dimensional Ultrasound Images for Chronic Lower Back Pain (cLBP) Assessment

Zixue Zeng [1,2], Matthew Cartier[1], Xiaoyan Zhao[1], Pengyu Chen[1], Xin Meng[1], Zhiyu Sheng[3], Maryam Satarpour[1], John M Cormack[3], Allison C. Bean[7], Ryan P. Nussbaum[7], Maya Maurer[4], Emily Landis-Walkenhorst[4], Kang Kim[1,3], Ajay D. Wasan[4,5,*], Jiantao Pu[1,2,6,*]

[1]Department of Bioengineering, University of Pittsburgh, Pittsburgh, PA 15213, USA

[2] Department of Radiology, University of Pittsburgh, Pittsburgh, PA 15213, USA

[3]Cardiology, Department of Medicine, University of Pittsburgh, PA 15213, USA

[4] Department of Anesthesiology & Perioperative Medicine, University of Pittsburgh, School of Medicine, PA 15213, USA

[5]Department of Psychiatry, University of Pittsburgh, School of Medicine, PA 15213, USA

[6] Department of Ophthalmology, University of Pittsburgh, Pittsburgh, PA 15213, USA

[7] Department of Physiatry, University of Pittsburgh School of Medicine, PA 15213, USA

* Indicates dual senior co-authors with equal contributions

**Acknowledgement:** This work is supported in part by research grants from the National Institutes of Health (NIH) (R61AT012282 and R01CA237277).

**Key Points**

1) **Detailed and Accurate Segmentation:** Six tissue layers in the lower back at the L4/L5 location are segmented on three-dimensional B-mode ultrasound images.
2) **Reduces Labeling Requirements:** Continuous image translation is used to generate inter-slice ultrasound images to reduce manual labeling efforts.

**Abbreviations**

cLBP: Chronic Lower Back Pain

CT: Computed Tomography

CSA: Cross-sectional Area

TrA: Transversus Abdominis

TLF: Thoracolumbar Fascia

MF: Multifidus

PSV: Peak Systolic Velocity

PSVR: PSV Ratio

3D: Three-dimensional

AI: Artificial Intelligence

RCA: Row-column Array

SFM: Superficial Fascial Membrane

DFM: Deep Fascial Membrane

MF: Multifidus

GAN: Generative Adversarial Network

IMP: Image-mask Pair

FID: Fréchet Inception Distance

IS: Inception Score

**Abstract**

**Background:** Available studies on chronic lower back pain (cLBP) typically focus on one or a few specific tissues rather than conducting a comprehensive layer-by-layer analysis. Since three-dimensional (3-D) images often contain hundreds of slices, manual annotation of these anatomical structures is both time-consuming and error-prone.

**Objectives:** We aim to develop and validate a novel approach called InterSliceBoost to enable the training of a segmentation model on a partially annotated dataset without compromising segmentation performance.

**Materials and Methods:** The architecture of InterSliceBoost includes two components: an inter-slice generator and a segmentation model. The generator utilizes residual block-based encoders to extract features from adjacent image-mask pairs (IMPs). Differential features are calculated and input into a decoder to generate inter-slice IMPs. The segmentation model is trained on partially annotated datasets (e.g., skipping 1, 2, 3, or 7 images) and the generated inter-slice IMPs. To validate the performance of InterSliceBoost, we utilized a dataset of 76 B-mode ultrasound scans acquired on 29 subjects enrolled in an ongoing cLBP study. An ultrasound operator annotated six anatomical layers across all image slices (n=18,986), including dermis, superficial fat, superficial fascial membrane (SFM), deep fat, deep fascial membrane (DFM), and muscle. The mean dice coefficient across all tissue layer was considered as primary outcome, and a paired sample t-tests with Bonferroni correction was conducted to assess the significance of performance differences.

**Results:** InterSliceBoost, trained on only 33% of the image slices, achieved a mean Dice coefficient of 80.84% across all six layers on the independent test set, with Dice coefficients of 73.48%, 61.11%, 81.87%, 95.74%, 83.52% and 88.74% for segmenting dermis, superficial fat, SFM, deep fat, DFM, and muscle. This performance is significantly higher than the conventional model trained on fully annotated images ($p<0.05$).

**Conclusion:** InterSliceBoost can effectively segment the six tissue layers depicted on 3-D B-model ultrasound images in settings with partial annotations.

**Clinical Relevance Statement**: InterSliceBoost enables tissue layer segmentation with fewer manual annotations. By comprehensively examining each tissue layer, this approach offers deeper insights into the structural and functional characteristics that may contribute to the onset and progression of cLBP.

# 1. Introduction

Chronic lower back pain (cLBP) is a complex and prevalent condition. It affects a large portion of the population and is one of the leading causes of disability worldwide [1]. It is estimated that over a quarter of adults in the United States experience lower back pain [2]. cLBP greatly diminishes the quality of life, causing persistent pain, sleep disturbances, and emotional distress [3-5]. Despite extensive research, reliable biomarkers closely associated with the presence of severity of cLBP remain elusive, and no optimal treatment has been identified [6].

Several clinical imaging modalities, including computed tomography (CT), magnetic resonance imaging (MRI), and ultrasound, have been utilized to investigate the underlying mechanism of cLBP. CT imaging can provide detailed views of bone and structural changes, making it effective for assessing alternations in the lumbar spine and surrounding muscles, such as a smaller cross-sectional area (CSA) of the multifidus muscle in LBP patients [7,8]. However, CT involves ionizing radiation and is less effective for evaluating soft tissues. MRI excels in detecting disc degeneration, herniation, Modic changes, and spinal stenosis in cLBP patients [9]. It offers high-resolution images of soft tissues without exposing patients to ionizing radiation; however, it is more expensive, less accessible, and less time-efficient than CT imaging. In contrast, while ultrasound imaging has lower image quality, it provides real-time tissue characterization without ionizing radiation [10], making it a valuable tool for studying cLBP, and particularly, pain related to soft tissues, such as myofascial pain. B-mode ultrasound studies have revealed associations between cLBP and various tissue abnormalities, such as reduced muscle activation in the transversus abdominis (TrA) muscle, evidenced by a small increase in thickness compared to healthy controls [11-16]. The thoracolumbar fascia (TLF) has shown increased thickness and echogenicity, correlating with pain intensity and decreased muscle function [17-22]. The multifidus (MF) muscle demonstrated distinct morphological and functional impairments, including asymmetrically smaller multifidus muscles and increased stiffness [23-29]. Doppler ultrasound has identified increased peak systolic velocity (PSV) as a potential factor in lumbar arteries of LBP patients, particularly at the L5 level, suggesting inflammatory hyperemia as a potential factor of cLBP pathology [30]. Additionally, a higher normalized artery-to-aortic PSV ratio (PSVR) was observed in a different study [31], indicating localized inflammation.

Most existing studies on cLBP focus on analyzing the entire image or a limited number of tissues, often neglecting a thorough layer-by-layer analysis from the dermis to the muscle. This narrow scope may overlook critical information about specific tissues, their interactions, and their collective impact on cLBP. Additionally, much of the current research relies primarily on two-dimensional (2D) ultrasound images, which may fail to capture the complex volumetric and detailed anatomical structures necessary for a thorough understanding. In contrast, three-dimensional (3D) ultrasound imaging has no such limitations but presents its own challenges. It generates a substantial number of images (often exceeding 150 slices) and is constrained by factors such as restricted penetration depth and susceptibility to image artifacts and noise, which can compromise image quality. These factors make manual annotation of tissue layers a time-intensive task. Consequently, developing an automated segmentation approach is critical for accurately identifying imaging layer-by-layer biomarkers associated with cLBP from 3-D ultrasound data.

A variety of techniques have been developed to address the challenge of training Artificial Intelligence (AI) models with limited annotated data, including few-shot learning (FSL) [32], sample-efficient learning (SEL), semi-supervised learning (SSL)[33], and Generative Adversarial Networks (GANs) [34]. FSL focuses on quickly generalizing to new tasks with minimal data, while SEL aims to maximize model performance using fewer training examples, often through data augmentation methods like geometric transformations (e.g., flipping, scaling) and intensity modifications (e.g., Gaussian noise) [35]. However, traditional augmentations may fail to capture real-world complexities, limiting generalization. SSL enhances performance by combining supervised learning on labeled data with unsupervised learning on a larger pool of unlabeled data. However, its effectiveness depends heavily on the quality of labeled data and assumptions about data structure, such as smoothness or clustering. GANs generate synthetic images to augment datasets, improving diversity, but the pretraining process and fixed synthetic outputs can restrict their utility[36]. Moreover, while GANs can enhance image quality, this does not always translate into improved segmentation model performance.

In this study, we present a novel method called InterSliceBoost that is designed to skip the annotation of a certain number of image slices. The underlying idea is to utilize generative AI to automatically produce intermediate images and annotations between the neighboring annotated slices. Unlike traditional autoencoder structures, InterSliceBoost employs a residual-block-based encoder to compute differential feature maps between adjacent slices. These feature maps are scaled using a warp ratio, allowing the generation of interslice images and masks at any desired location between adjacent slices. We expect that this method will not only enhance segmentation performance by increasing image but also reduce the burden of manual annotation, facilitating more accurate tissue layer feature extraction as downsteam tasks. A detailed description of the methods and the experimental results follows.

## 2. Materials and Methods

**2.1 Study Dataset**

We collected and annotated a dataset of 76 B-mode ultrasound scans from 29 subjects enrolled in an ongoing study on cLBP (Table 1). Of these participants, 21 have been diagnosed with cLBP. To ensure consistent image acquisition, participants lay prone on an exam table with the ultrasound array transducer placed laterally to the midline at the L3-L4 vertebral interspace (Figure 1, Part 1) over the MF and erector spinae (ES) muscles, respectively. A row-column array (RCA) transducer (RC6gV, Vermon) with a center frequency of 6 MHz and an active aperture of 25.6 mm by 25.6 mm connected to the Vantage 256 ultrasound system (Verasonics Inc., WA, USA) is used to capture 3D volumetric ultrasound data. To improve the image quality, the image acquisition sequence is programmed to utilize the synthetic aperture approach[37], which is, for each image frame, to individually transmit from each single element followed by synchronously receiving at all the elements. This approach enables the dynamic focus on each spatial location during beamforming and image reconstruction. Temporal compounding is also applied using 3 consecutive frames to enhance the image signal-to-noise ratio. Scanning is conducted on both the right and left sides of the back. On each side, the scan is performed at 2 locations, over the MF and ES muscles, respectively. The scan is repeated 3 times at each location. This results in 12 B-mode scans per participant. In total, 348 B-mode scans were obtained. For the purpose of algorithm development and validation, we randomly selected 76 scans acquired from different locations. Ultrasound examinations were performed by board-certified physiatrists or pain specialists with advanced training in diagnostic and interventional musculoskeletal ultrasonography, with 9 (ACB), 7 (RPN), and 25 (ADW) years of clinical experience.

An ultrasound operator (Ms. Zhao) meticulously outlined six anatomical layers, including dermis, superficial fat, superficial fascial membrane (SFM), deep fat, deep fascial membrane (DFM), and MF muscle (Figure 1, Parts 2 & 3). The annotated dataset was divided into two subsets at the patient level: Subset A (patients =19) and Subset B (patients=10) (Figure 4 section 1). Subset A is used to train the segmentation model, the deblurring model, and the inter-slice generation model, while Subset B is used as an independent test set to evaluate the transferability of the inter-slice generation model to new patient images. Subset A is further split into training, internal validation, and independent test sets at the patient level, with the training set containing 10,670 images (patients=16), the internal validation set containing 508 images (patients=2), and the independent test set containing 976 images (patients=1).

**Table 1.** Descriptive statistics information for demographic variables and LBP status.

| N = 29 | Summary |
|---|---|
| Demographic variable | |
| Age | 47.11(19.34) |
| Female | 18(62.07%) |
| Hispanic or Latino | 1(3.44%) |
| Black or African American | 2(6.90%) |
| White | 26(89.65%) |
| Height (in inches) | 66.38(3.80) |
| Weight (in pounds) | 179.83(36.02) |
| cLBP status | |
| Positive | 21(72.41%) |

Summary statistics are reported as mean (standard deviation) for continuous measurements and number of individuals (percentages) for categorical measurements.

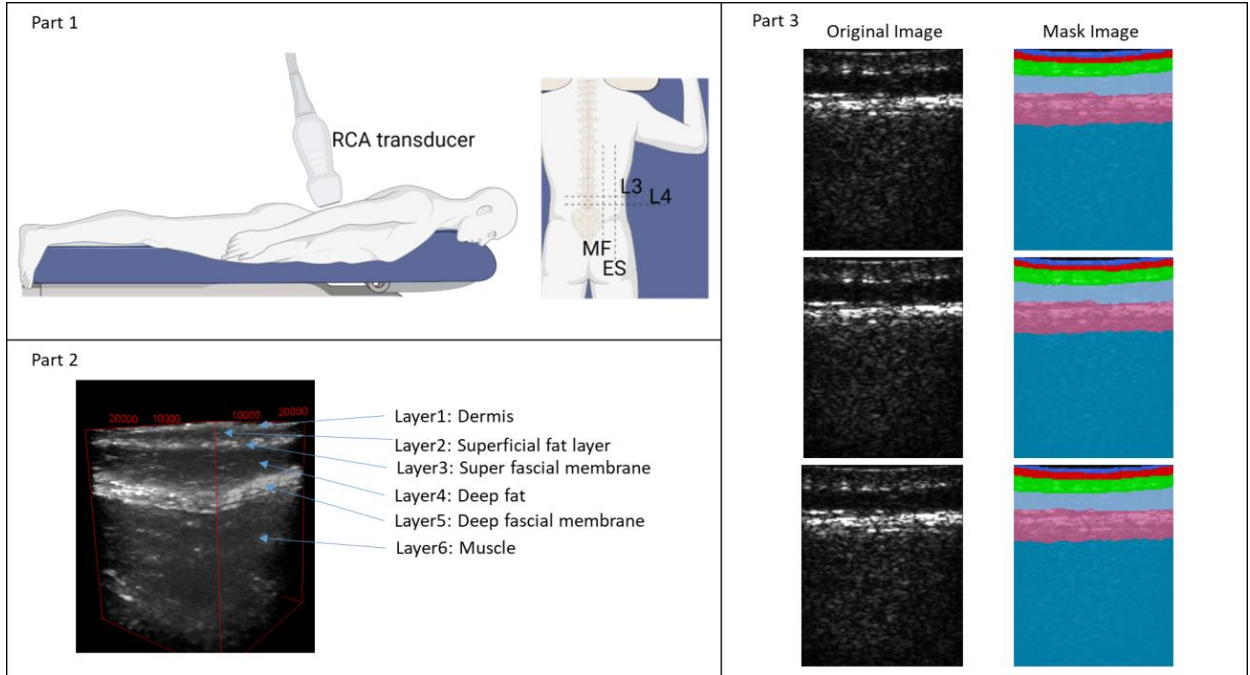

**Figure 1.** 3-D ultrasound B-mode image acquisition and annotations. Part 1: ultrasound image acquisition. Part 2: A 3-D ultrasound B-mode image depicting six tissue layers. Part 3: Original B-mode images along the coronal or sagittal directions paired with their corresponding annotations, where each color represents a specific tissue layer in Part 2.

## 2.2 An inter-slice Image Generator

The inter-slice image generator is based on GAN. In architecture, it consists of a residual block-based encoder, a UNet-style decoder, a PatchGAN discriminator, and an optional deblurring model (Fig. 2). The input tensor of the encoder is formed by the image and the corresponding mask, namely two-channel image-mask pair (IMP). Both left-side IMP and right-side IMP are processed by the encoder separately to generate two sets of feature maps. The difference calculation mechanism produces $F_{difference}$ based on $(F_{right} - F_{left}) \times$ ratio, where the ratio reflects the position of the inter-slice IMP. The decoder then takes $F_{left}$ and $F_{difference}$ as inputs to reconstruct the inter-slice image and mask using the feature maps. The PatchGAN discriminator classifies the generated inter-slice image as either real or fake. This architecture combines residual learning with a UNet-style design for inter-slice image generation, aiming to offer precise control over the slice position. The deblurring model is optional and used to enhance the quality of the generated inter-slice IMPs by refining image details. This model only processes the original images associated with each IMP, without incorporating the corresponding masks.

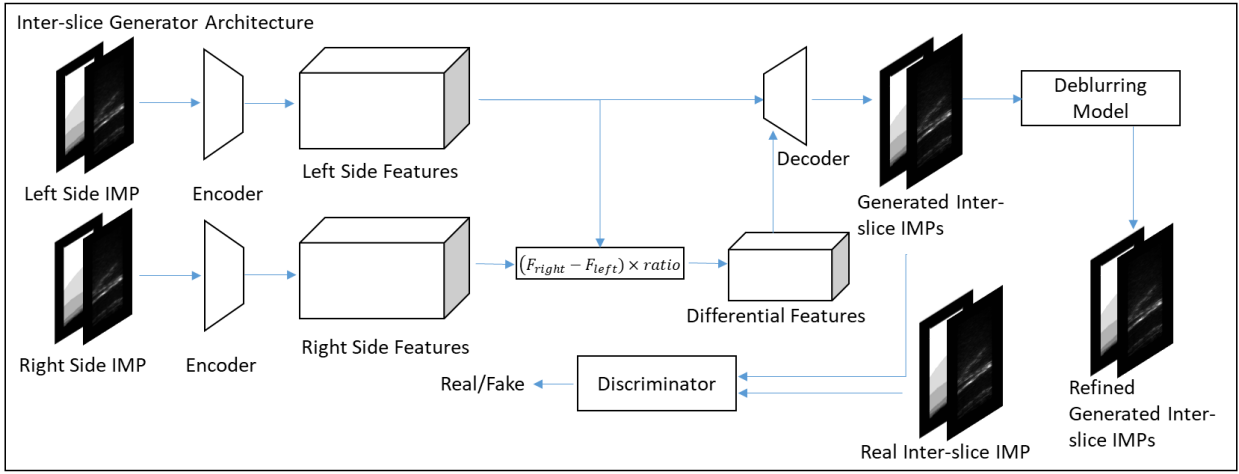

**Figure 2.** Inter-slice generator architecture. The left-side and right-side image-mask-pairs (IMPs) are processed separately by the encoder to produce corresponding feature maps. The ratio parameter is used to specify the location of the generated inter-slice within the left and right IMPs. The decoder then receives the differential feature and the left side IMP's feature as inputs to generate the inter-slice IMP based on the specified ratio. The discriminator evaluates the generated inter-slice IMP against the real inter-slice IMP.

**2.3 Backbone Model Architecture**

The encoder shown in the upper section of Figure 3 utilizes residual blocks to facilitate efficient gradient flow and enable deeper network training. It begins with an initial convolutional layer to reduce spatial dimensions and increase feature depth, followed by a series of residual blocks that progressively down-sample the input to capture hierarchical features. The decoder shown in the lower section of Figure 2 employs 2-D transposed convolutions to reconstruct high-dimensional feature representations into image-level outputs. Following the pix2pix framework[38], a PatchGAN discriminator is incorporated to evaluate the realism of the generated images at the patch level. The discriminator comprises multiple convolutional layers that gradually extract features from input images and generate a patch-wise label map to assess the authenticity of each patch within the image. Similarly, the deblurring model adapts an architecture akin to the pix2pix model, utilizing a UNet for image generation and a PatchGAN discriminator to evaluate the quality of the generated images.

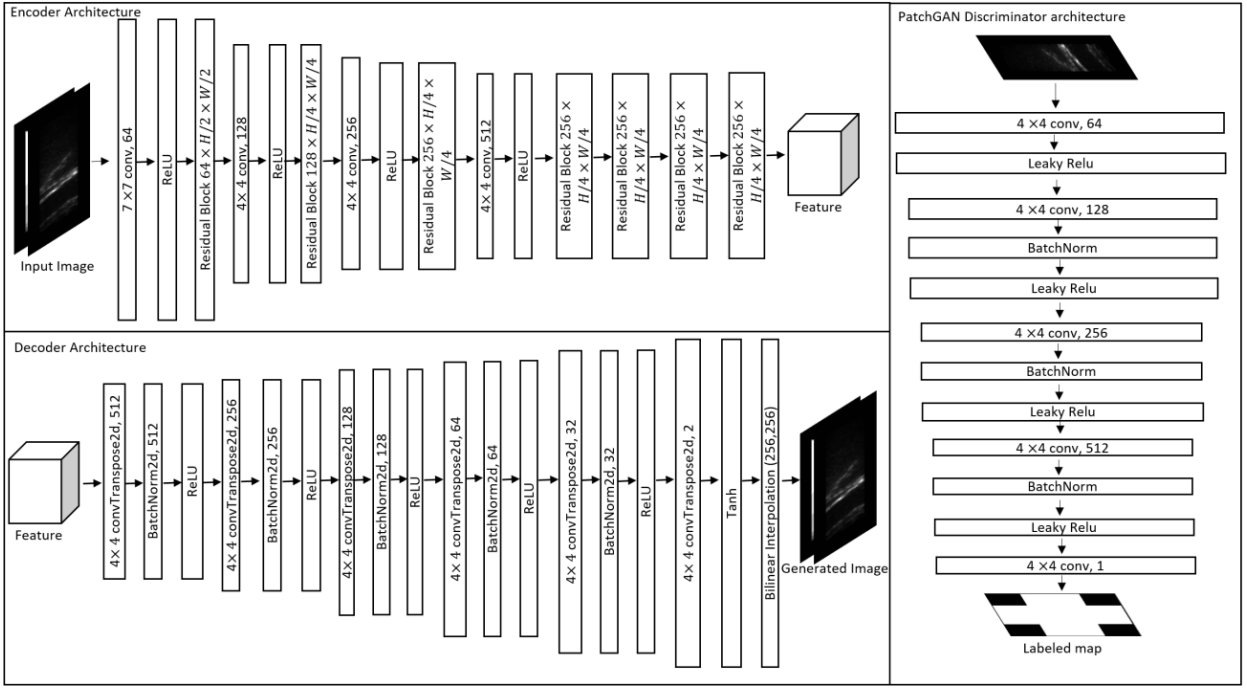

**Figure 3.** Backbone Model Architecture. The encoder (upper section, left) utilizes multiple residual blocks to progressively capture high dimensional patterns. The decoder (lower section, left) reconstructs the image-mask pairs (IMPs) from these high-level representations using transposed convolution layers. The PatchGAN discriminator (right) utilizes a series of convolutional layers to analyze the generated outputs and produces a patch-wise label map to assess their authenticity.

**2.4 Loss Function**

Dice coefficient is used as the loss function for the segmentation model. For the inter-slice image generator model, a composite loss function is used to train both the generator and the discriminator. The generator's loss combines L1 loss across different ratios:

$$Loss_G = \lambda_{L1}(loss_{L1_0} + loss_{L1_{0.5}} + loss_{L1_1}) + \lambda_{adv}(loss_{GAN_0} + loss_{GAN_{0.5}} + loss_{GAN_1}) \quad (1)$$

where $loss_{L1_0}$, $loss_{L1_{0.5}}$, and $loss_{L1_1}$ represent the L1 losses for the ratios 0, 0.5, and 1, respectively. The functions $loss_{GAN_0}$, $loss_{GAN_{0.5}}$ and $loss_{GAN_1}$ represent the GAN losses for these ratios. $\lambda_{L1}$ and $\lambda_{adv}$ denote L1 loss rate and discriminator loss rate. Following prior research[38], $\lambda_{L1}$ and $\lambda_{adv}$ are set to be 100 and 1 respectively. The L1 loss is defined as:

$$L1\ Loss = \frac{1}{N}\sum_{i=1}^{N}\|G(x_i) - y_i\|_1 \quad (2)$$

where N is the number of images in each batch. $G(x_i)$ represents the generated image, and $y_i$ refers to the corresponding ground truth images. The GAN loss in the generator's loss function is defined as:

$$loss_{GAN} = -E\left(\log\left(D(G(x))\right)\right) \tag{3}$$

where $D(G(x))$ represents the discriminator's probability estimate that the generated image $G(x)$ is real.

The discriminator's loss combines the real and fake losses across different ratios:

$$Loss_D = 0.25 \times \left(loss_{D_{real}} + loss_{D_{fake_0}} + loss_{D_{fake_{0.5}}} + loss_{D_{fake_1}}\right) \tag{4}$$

where:

$$loss_{D_{real}} = -E(\log(D(y))) \tag{5}$$

$$loss_{D_{fake}} = -E\left(\log\left(1 - D(G(x))\right)\right) \tag{6}$$

where $loss_{D_{fake_0}}$, $loss_{D_{fake_{0.5}}}$, and $loss_{D_{fake_1}}$ represent the discriminator loss on fake images corresponding to the ratios 0, 0.5, and 1, respectively. $G(x)$ denotes the generated image, and $y$ refers to the ground truth images.

For the deblurring model, the generator's loss is a combination of GAN loss and L1 loss

$$Loss_G = \lambda_{L1} loss_{L1} + \lambda_{adv} loss_{GAN} \tag{7}$$

Similar to Eq. (1), $\lambda_{L1}$ and $\lambda_{adv}$ are set to 100 and 1, respectively. The discriminator loss of the deblurring model is the average of the real and fake losses:

$$Loss_D = 0.5 \times \left(loss_{D_{real}} + loss_{D_{fake}}\right) \tag{8}$$

where $loss_{D_{real}}$ denotes the discriminator's loss on real images, while $loss_{D_{fake}}$ represents the discriminator's loss on fake images.

## 2.5 Training Protocol and Experiment Setting

After collecting and preprocessing the ultrasound images, the dataset was split at the patient level to prevent data leakage. The dataset was divided into two subsets: Subset A and Subset B (Figure 4, Part 2). Subset A was used to train the segmentation model, deblurring model, and inter-slice generation model, while Subset B was reserved for evaluating the generalizability of the inter-slice generation model to new patient images. Subset A is further divided into training, internal validation, and independent test sets, with 10,670 images in the training set, 508 images in the internal validation set, and 976 images in the independent test set. Subset B, used for generalizability evaluation, contains 6,832 images. Each image in the datasets was paired with its corresponding mask. To assess the impact of varying annotation proportion on segmentation performance, the original study dataset was processed into partially annotated subsets across four different settings, which were used to train the inter-slice image generator model and the deblurring model:

1) Setting 1 (50% labeled): For each pair of adjacent images, the second image is not annotated.
2) Setting 2 (33% labeled): For each group of three adjacent images, the last two images are not annotated.
3) Setting 3 (25% labeled): For each group of four adjacent images, the last three images are not annotated.
4) Setting 4 (12% labeled): For each group of eight adjacent images, the last seven images are not annotated.

This approach generates partially annotated datasets with fewer annotated images, allowing the inter-slice generation model to generate IMPs to fill in the gaps, resulting in the same number of samples as the original dataset. Below is an overview of our experimental workflow (Figure. 4):

Step 1: Various segmentation model were trained on the training data from Subset A, and evaluated based on internal validation set performance. To ensure high-quality biomarker extraction for all tissue layers, the best-performing model was selected based on average dice coefficient across tissue layers for future use[39]. (Part 3 in Figure 4)

Step 2: The inter-slice image generator model was trained on the partially annotated datasets and used to generate two additional datasets: a partially annotated dataset for training the deblurring model and an interpolated dataset for inter-slice augmentation of the segmentation model. The deblurring model was trained on the generated partially annotated dataset, using the original partially annotated dataset as the ground truth. After training, the deblurring model was applied to post-process the interpolated dataset. (Part 4-5 in Figure 4)

Step 3: The segmentation model selected from Step 1 was then trained on the partially annotated datasets and the interpolated datasets formed in Step 2. Additionally, the model was trained on the inter-slice augmented dataset where images have been post-processed by the deblurring model. (Part 6 in Figure 4). These steps were repeated for all four settings, and the performance of the model across these different training settings was compared. Additionally, we compared our InterSliceBoost method with several other methods, including traditional image augmentation, bilinear interpolation, frame-interpolation[40], GAN-based augmentation, VAE-based augmentation[41], and AdvChain[42] in setting 1 (12% labeled). GAN-based augmentation involves training a GAN model for reconstruction purposes and then using the pre-trained generator to create a combined dataset ($D_{original}, D_{reconstructed}$). This joint dataset is subsequently used to train the segmentation model.

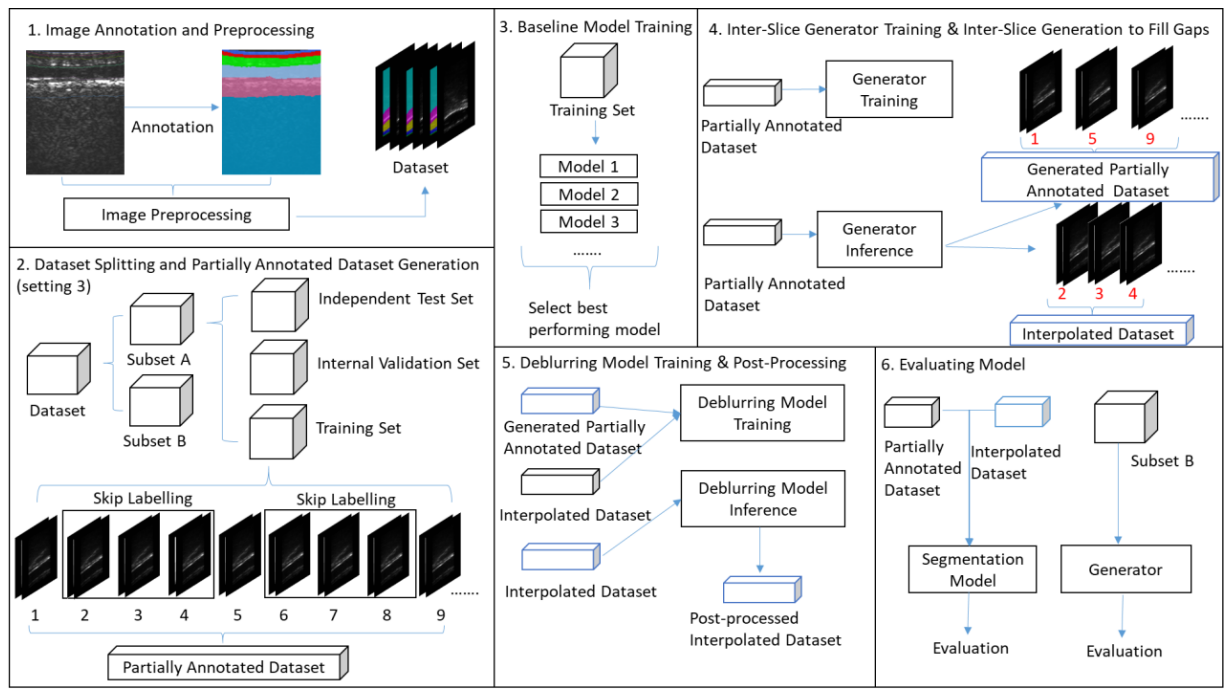

**Figure 4.** The overall workflow for training the segmentation models using inter-slice augmentation and deblurring techniques. In Setting 3, only the first image in each group of four was retained, creating a partially annotated dataset. Part 1: images are annotated and preprocessed. Part 2: the study dataset is split into Subset A and Subset B. Subset A is split into training, internal validation, and independent test sets. Some image slices in the training set are skipped, leading to the partially annotated dataset. Part 3: train segmentation models and select the best model. Part 4: train inter-slice generator to generate interpolated dataset based on the partially annotated dataset. Part 5: train the deblurring model and perform post-processing. Part 6: evaluate the performance of the segmentation model.

The training strategy for the inter-slice generation model involves an approach where the generator is conditioned on different ratios that represent spatial positions within adjacent slices of the partially annotated datasets (Figure 5). During training, the ratio parameters were set to 0, 0.5, and 1, with 0 corresponding to the left-side IMP, 1 to the right-side IMP, and 0.5 to the middle IMP. The generator takes the IMPs at ratios 0 and 1 as inputs and generates IMPs at ratios 0, 0.5, and 1 as outputs. The generated outputs are concurrently used to train the discriminator, which evaluates the authenticity of the generated IMPs by comparing them to real IMPs. During the inference stage, custom ratios between 0 and 1 are set. The pretrained inter-slice generator takes two adjacent IMPs as input and generates inter-slice IMPs to fill the gaps between the existing slices.

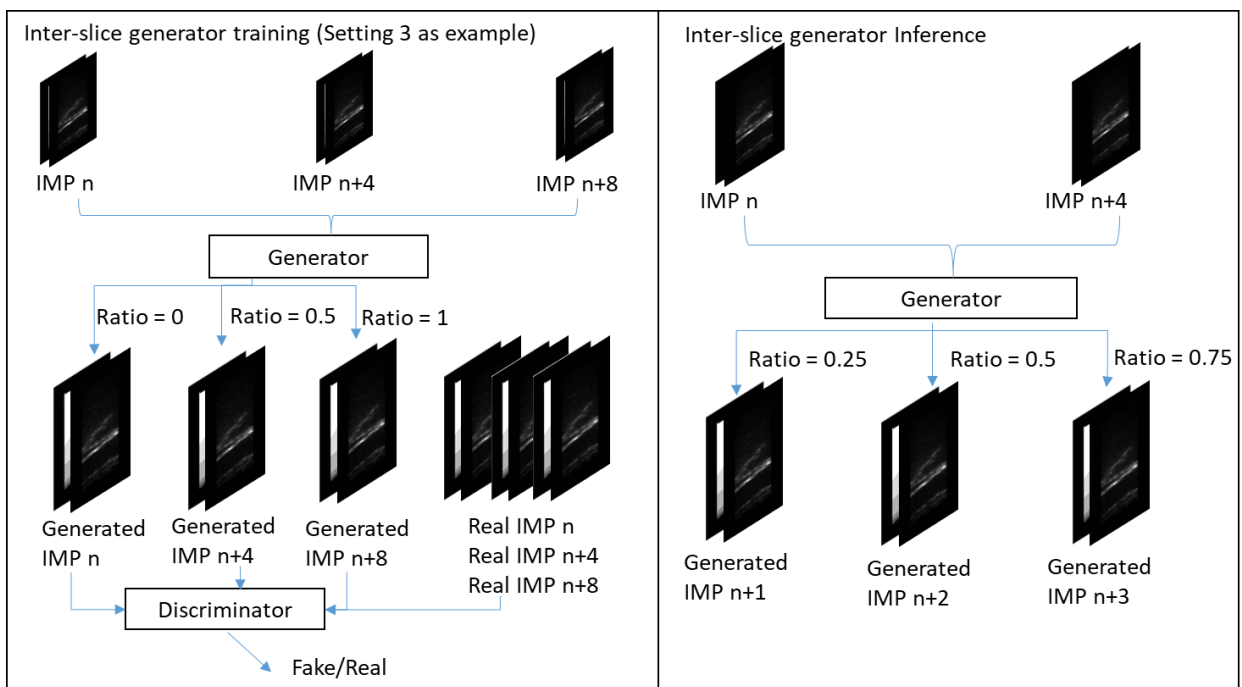

**Figure 5.** Inter-slice generator training and inference workflow in setting 3. Training (left): The IMP ratios are set to be 0, 0.5, and 1. The generated slices at these ratios are input into the discriminator for adversarial training. Inference (right): The IMP ratios are set to be 0.25, 0.5, and 0.75, to generate slices that fill the gaps between existing slices.

For image augmentation, we used the "ImgAug" library to apply a range of geometric and intensity transformations. The augmentation pipeline includes random cropping, padding, flipping, affine transformations (e.g., scaling, translation, rotation, shearing), and adjustments to image quality, such as adding noise, changing contrast, sharpening, and blurring.

For the segmentation model, training was performed using the Adam optimizer with a learning rate of 0.0001 and a decay rate of 0.1. The Dice coefficient was used as the loss function. The maximum number of epochs was 50 with a patience of 5 epochs. The inter-slice image generator model was trained with the Adam optimizer at a learning rate of 0.0002 and with exponential decay rates $\beta$ set to (0.5, 0.999). The model with the highest Fréchet Inception Distance (FID) on the validation dataset was selected. The deblurring model also used the Adam optimizer with a learning rate of 0.0002 and exponential decay rate $\beta$ parameters set to (0.5, 0.999). In the above training, early stop mechanism was used, namely the training would stop if the validation performance did not improve for 5 consecutive epochs. All experiments were implemented using PyTorch version 2.3.0 and CUDA version 12.1 on NVIDIA GTX 3090 GPU.

## 2.6 Performance evaluation

The performance of the models in this study was evaluated using a set of metrics tailored to specific tasks, including segmentation, inter-slice image generator, and deblurring. The segmentation model was evaluated using the dice coefficient per tissue layer and the mean dice coefficient across all tissue layers, with a higher dice coefficient indicating a better overlap between predicted and ground truth masks. For exploratory analyses involving tissue-layer comparisons across different models, we used paired sample t-tests with a 0.05 significance threshold, without applying multiple-comparisons corrections. In contrast, to maintain control of the family-wise error rate for our primary outcome (the mean Dice coefficient), we employed Bonferroni correction for multiple comparisons among the models, following the formula:

$$\alpha_{adjusted} = \frac{\alpha}{m} \tag{5}$$

where $m$ is the number of comparisons and $\alpha$ is significance level.

The inter-slice generator model and the deblurring model were evaluated using the FID. FID measures the quality of generated images by comparing their statistical properties to those of real images. It is calculated by passing the images through a pre-trained inception network and computing the mean and covariance of the feature activations. The FID score is given by

$$\text{FID} = |\mu_r - \mu_g|^2 + Tr\left(\Sigma_r + \Sigma_g - 2\sqrt{\Sigma_r \Sigma_g}\right), \tag{6}$$

where $\mu_r$ and $\Sigma_r$ are the mean and covariance of the real images' features, and $\mu_g$ and $\Sigma_g$ are the mean and covariance of the generated images' features. $Tr()$ stands for the trace of matrix. A lower FID score indicates better quality and higher similarity between the generated and real images.

The inter-slice generation model was evaluated using the Inception Score (IS), a metric that measures both the resemblance between the real and generated images and the diversity of the generated images. This score is particularly useful for assessing generative models such as GANs [43]. The IS was calculated using a pre-trained Inceptionv3 network to obtain the conditional label distribution $p(y|x)$ for each generated image $x$, where $y$ is the label predicted by the network. The IS is given by

$$\text{IS} = exp\left(E_x[D_{KL}(p(y|x)||p(y))]\right), \tag{6}$$

where $D_{KL}$ denotes the Kullback-Leibler divergence between the conditional distribution $p(y|x)$ and the marginal distribution $p(y)$.

To assess the impact of inter-slice augmentation and compare it with other state-of-the-art approaches, we computed Cohen's effect size $d$ using the following fomula[44]:

$$d = \frac{M_2 - M_1}{\sqrt{\frac{SD_1^2 + SD_2^2}{2}}} \tag{7}$$

where $M_1$ and $M_2$ are the means of the two sample groups. $SD_1^2$ and $SD_2^2$ are the standard deviations of two sample groups. Effect sizes are generally classified as follows: d < 0.2 indicates a very small effect, 0.2 ⩽ d < 0.5 represents a small effect, 0.5 ⩽ d < 0.8 corresponds to a medium effect, and d ⩾ 0.8 denotes a large effect size.

## 3. Results

### 3.1 Baseline Model Results

The performance of various segmentation models was evaluated on the internal validation set, with results summarized in Table 2. Among the baseline models, Deeplabv3plus achieved the highest mean Dice score of 85.15%, significantly outperforming other models, and achieving a 12.17% higher dice score for the most challenging tissue layer (i.e. dermis) while maintaining comparable performance for SFM, Deep Fat, DFM, and Muscle layers. Consequently, Deeplabv3plus was selected for subsequent experiments to further evaluate and enhance segmentation performance through inter-slice augmentation and other techniques.

**Table 2.** Baseline model performance metrics (mean Dice coefficient) on the internal validation set of Subset A (n=2).

| Model Name | Dermis↑ | Superficial Fat Layer↑ | SFM↑ | Deep Fat↑ | DFM↑ | Muscle↑ | Overall↑ |
|---|---|---|---|---|---|---|---|
| Dynnet | 71.24(0.12) | 70.73(0.13) | 80.45(0.08) | 87.59(0.06) | 74.11(0.09) | 91.94(0.03) | 79.34(0.08) |
| Segresnet | 65.35(0.18) | 68.61(0.17) | 85.91(0.04) | 94.87(0.02) | 83.84(0.09) | 93.26(0.02) | 81.97(0.11) |
| UNet | 67.62(0.16) | 66.27(0.18) | 81.69(0.06) | 85.58(0.09) | 80.89(0.08) | 90.55(0.04) | 78.77(0.05) |
| Attention-UNet | 66.97(0.17) | 73.07(0.14) | 88.09(0.04) | 92.52(0.04) | 81.34(0.09) | 92.95(0.02) | 82.49(0.10) |
| VNet | 64.60(0.16) | 76.48(0.10) | 89.24(0.04) | 95.40(0.01) | 85.64(0.07) | 94.38(0.03) | 84.29(0.09) |
| **Deeplabv3plus** | **76.77**(0.11) | 71.03(0.10) | 88.93(0.04) | 95.11(0.02) | 85.02(0.08) | 94.02(0.03) | **85.15**(0.089) |

SFM = Super Fascial Membrane, DFM = Deep Fascial Membrane. Values are presented as mean Dice coefficients in percentage, with the standard deviation in parentheses. Bold values indicate instances where Deeplabv3plus significantly outperformed other methods ($p<0.05$ for tissue-layer comparisons and $p<0.05/5$ for the mean Dice coefficient, paired t-test). ↑ indicates a higher value is preferable.

### 3.2 Comparison with other state-of-the-art methods

Table 3 presents the performance of our InterSliceBoost model alongside several other state-of-the-art data augmentation methods on the independent test set in Setting 4 (12% labeled). InterSliceBoost significantly outperformed other methods, both with image augmentation and without augmentation.

**Table 3.** Comparison of segmentation model performance between our method and existing methods on the independent test set in Setting 4 (12% labeled), evaluated using the Dice coefficient (Standard deviation).

| Model Name | Dermis↑ | Superficial Fat Layer↑ | SFM↑ | Deep Fat↑ | DFM↑ | Muscle↑ | Overall↑ | Effect size↑ |
|---|---|---|---|---|---|---|---|---|
| Baseline | 48.33(0.21) | 58.36(0.14) | 82.65(0.10) | 95.98(0.02) | 85.72(0.07) | 90.31(0.05) | 76.91(0.17) | - |
| Baseline w aug | 48.54(0.21) | 59.02(0.14) | 82.29(0.09) | 95.92(0.02) | 87.30(0.06) | 91.60(0.05) | 77.44(0.17) | 0.031 |
| Bilinear Interpolation | 59.53(0.16) | 40.22(0.21) | 71.61(0.18) | 94.38(0.02) | 87.02(0.06) | 91.92(0.05) | 74.11(0.19) | -0.155 |
| Frame Interpolation[45] | 0.00(0.00) | 56.37(0.14) | 79.26(0.09) | 95.59(0.01) | 87.98(0.06) | 92.79(0.03) | 68.67(0.33) | -0.314 |
| GAN-based Augmentation | 50.16(0.20) | 57.64(0.13) | 81.99(0.09) | 95.89(0.02) | 88.08(0.06) | 91.99(0.05) | 77.63(0.17) | 0.042 |
| VAE-based Augmentation[41] | 46.35(0.22) | 57.73(0.15) | 82.87(0.09) | 95.94(0.02) | 83.69(0.08) | 88.74(0.05) | 75.87(0.17) | -0.061 |
| AdvChain[42] | 45.87(0.22) | 58.18(0.13) | 81.57(0.09) | 95.85(0.02) | 83.79(0.09) | 88.90(0.07) | 75.69(0.17) | -0.072 |
| InterSliceBoost w.o. aug | **62.69(0.11)** | 59.46(0.13) | 82.44(0.08) | 95.94(0.02) | 84.75(0.07) | 89.96(0.006) | **79.21(0.13)** | **0.152** |
| InterSliceBoost | **61.29(0.13)** | 60.21(0.13) | **83.89(0.08)** | 95.88(0.02) | 86.05(0.08) | 90.62(0.06) | **79.66(0.14)** | **0.177** |

Note: "Baseline" refers to the model trained on the partially annotated dataset (12% labeled). Bold text indicates the Interslice Boost performance, which significantly outperformed all other methods ($p<0.05$

for tissue-layer comparisons and p<0.05/7 for the mean Dice coefficient, paired t-test). 'aug' denotes the use of image augmentation. ↑ indicates a higher value is preferable.

### 3.3 Inter-slice Augmentation

Table 4 summarizes the mean Dice coefficient for each tissue layer and the overall mean across each setting, with and without inter-slice augmentation. The Deeplabv3plus model achieved Dice coefficients of 83.02%, 95.94%, 87.98%, and 93.00% for SFM, Deep Fat, DFM, and muscle in default setting, respectively. The inter-slice augmentation mechanism significantly improved the model's performance across all settings (p<0.05). In settings 1 and 2, where the images are partially annotated, the model performed even better compared to the default setting (p<0.05). In setting 3 and 4, however, the Deeplabv3plus model showed a slight decrease in performance compared to the default setting. Specifically in setting 4, the InterSliceBoost model achieved a Dice coefficient, only 0.77% lower than the fully supervised model, despite using only 12% of the labeled dataset. The segmentation model experienced a minimal and statistically insignificant decrease of 0.33% in performance when trained on IMPs that were post-processed by a deblurring model (p<0.05). Figure 6 shows the prediction results of the deeplabv3plus model, the inter-slice generator model, and the deblurring model. The Deeplabv3plus model, enhanced with our proposed inter-slice augmentation, achieved the most accurate tissue segmentation across all layers. Additionally, the inter-slice generator produced images that closely resembled the ground truth.

**Table 4.** Comparison of model performance comparison with inter-slice augmentation across different settings by the Dice coefficient (standard deviation).

| Setting (Labeling Proportions) | Inter-slice Augmentation | Dermis ↑ | Superficial Fat layer↑ | SFM↑ | Deep Fat↑ | DFM↑ | Muscle↑ | Overall↑ | Performance Change↑ |
|---|---|---|---|---|---|---|---|---|---|
| Fully Supervised | No | 62.28(0.10) | 60.42(0.12) | 83.02(0.09) | 95.94(0.02) | 87.98(0.06) | 93.00(0.04) | 80.43(0.14) | |
| Setting 1 (50%) | No | 61.04(0.12) | 61.67(0.12) | 82.53(0.09) | 95.88(0.02) | 87.44(0.07) | 91.85(0.05) | 80.06(0.14) | |
| Setting 1 (50%) | Yes | 71.70(0.13) | 60.19(0.13) | 81.36(0.09) | 95.95(0.02) | 87.61(0.07) | 91.88(0.05) | **81.45(0.12)** | **+1.39 (<0.01)** |
| Setting 2 (33%) | No | 65.40(0.10) | 57.52(0.13) | 82.48(0.09) | 95.85(0.02) | 88.05(0.06) | 92.42(0.05) | 80.28(0.14) | |
| Setting 2 (33%) | Yes | 73.48(0.13) | 61.11(0.13) | 81.87(0.09) | 95.74(0.02) | 83.52(0.09) | 88.74(0.07) | 80.84(0.11) | **+0.58 (<0.01)** |
| Setting 3 (25%) | No | 50.19(0.21) | 58.65(0.12) | 82.85(0.09) | 95.96(0.02) | 86.85(0.06) | 91.68(0.05) | 77.70(0.17) | |
| Setting 3 (25%) | Yes | 61.72(0.13) | 59.99(0.13) | 82.69(0.09) | 96.09(0.02) | 87.10(0.06) | 91.85(0.04) | 79.91(0.14) | **+2.21 (<0.01)** |
| Setting 4 (12%) | No | 48.54(0.21) | 59.02(0.14) | 82.29(0.09) | 95.92(0.02) | 87.30(0.06) | 91.60(0.05) | 77.44(0.17) | |
| Setting 4 (12%) | Yes | 61.29(0.13) | 60.21(0.13) | 83.89(0.08) | 95.88(0.02) | 86.05(0.08) | 90.62(0.06) | 79.66(0.14) | **+2.22 (<0.01)** |
| Setting 4 (12%) (Deblurred) | Yes | 49.42(0.22) | 56.43(0.15) | 83.61(0.09) | 96.02(0.02) | 86.07(0.08) | 91.13(0.06) | 77.11(0.17) | -0.33 (0.07) |

Note: "Fully supervised" refers to the model trained on the fully annotated dataset. Inter-slice augmentation indicates that the segmentation model was trained on a combined dataset of partially annotated datasets and the generated inter-slice IMPs. Performance changes are presented as percentages (p-value). Bold value in the overall column indicates settings with Dice coefficient significantly higher than the default setting, as determined by paired sample t-test (p<0.05). Bold values in performance change denote significant changes in the Dice coefficient compared with not using inter-slice augmentation, as determined by a paired sample t-test (p<0.05). Setting 4

(Deblurred) indicates that the inter-slice IMPs were post-processed by our deblurring model. ↑ indicates a higher value is preferable.

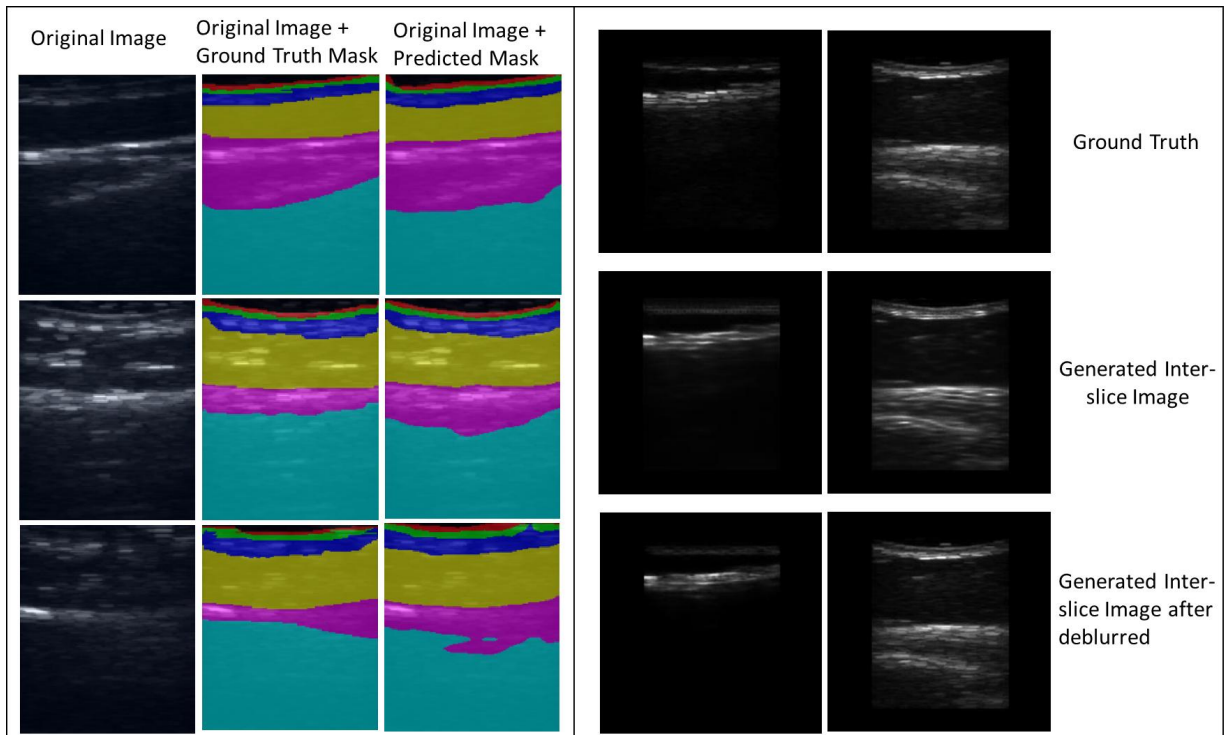

**Figure 6. Left:** Original images with both ground truth mask and predicted masks generated by the Deeplabv3plus model enhanced with our proposed inter-slice augmentation method. **Right:** Ground truth inter-slice images and generated inter-slice images, with and without deblurring.

The performance of the inter-slice generator model was evaluated on the validation set of Subset A and Subset B, an independent testing dataset, to assess its generalizability. The results in Table 5 show that the inter-slice generator model maintains stable performance when applied to the independent Subset B.

**Table 5.** Performance of the inter-slice generator on the validation set of Subset A and Subset B

| Setting where inter-slice generator trained on | Dataset | FID↓ | IS↑ |
| --- | --- | --- | --- |
| Setting 1 | Subset A validation | 232.23(63.32) | 1.17(0.04) |
| Setting 1 | Subset B | 166.78 (43.54) | 1.27 (0.07) |
| Setting 2 | Subset A validation | 328.94(113.31) | 1.19(0.04) |
| Setting 2 | Subset B | 190.30 (44.67) | 1.27 (0.06) |
| Setting 3 | Subset A validation | 229.59(73.95) | 1.15(0.06) |
| Setting 3 | Subset B | 160.30 (33.84) | 1.24 (0.07) |
| Setting 4 | Subset A validation | 224.63(41.73) | 1.09(0.05) |
| Setting 4 | Subset B | 151.82 (32.52) | 1.22 (0.06) |

Note: FID = Fréchet Inception Distance, IS = Inception Score. Subset B is the independent test dataset used to evaluate the performance of the inter-slice image generator. Performance metrics are presented as the mean (standard deviation). ↑ indicates that a higher value is preferable while ↓ indicates a lower value is preferable.

## 4. Discussion

We introduced a novel method, InterSliceBoost, a continuous deep learning-based image translation approach designed to segment tissue layers on B-model ultrasound images while reducing manual labeling efforts. By utilizing available partial annotation, the inter-slice generator can produce both images and annotations for all six tissue layers at any specified position between adjacent slices (Figure 2). The goal is to enhance data diversity and reduce manual annotations, ultimately improving the quality of image segmentation. Our experiments confirmed the feasibility of the solution: by reduction of labeling efforts by 67%, the model achieved a 0.41% higher dice coefficient compared to a fully annotated dataset (Table 4).

We designed the inter-slice generator architecture based primarily on the concept of continuous image translation. Unlike traditional methods, which typically generate inter-slice images based on bilinear interpolation between two adjacent images [46], our approach employs an encoder-decoder architecture combined with a PatchGAN discriminator (Figure 3). There are three reasons for this choice. First, the traditional bilinear interpolation method assumes linear changes between adjacent images. However, inter-slice shapes and features may follow complex, non-linear patterns that bilinear interpolation could fail to capture. In contrast, the deep learning-based generative model may capture the underlying non-linear variations. Second, ultrasound images are often noisy, and bilinear interpolation can inadvertently interpolate noisy pixels in a linear fashion. In contrast, the GAN-based model, by incorporating a PatchGAN discriminator, penalizes images with unrealistic noise artifacts, leading to the generation of more realistic and detailed inter-slice images. Finally, bilinear interpolation only uses information from two adjacent images, which increases the risk of overfitting. Our approach, however, generates new and diverse images by learning the images and their relationship, thereby enhancing the model generalization capabilities. The results from Setting 4 demonstrate the effectiveness of our method, with inter-slice augmentation yielding a higher effect size of 0.177 and a significantly improved Dice coefficient of 0.7966 (Table 3). These results significantly outperform bilinear interpolation and other state-of-the-art augmentation methods, further validating our choice to employ the inter-slice generator model.

The performance comparison across different settings (Table 4) shows that inter-slice augmentation significantly improves the model performance, regardless of the proportion of labeled data. By incorporating more inter-slice images and their corresponding annotations, the dataset gains increased diversity, leading to a more robust and reliable model. However, we observe that in Settings 3 and 4, where the proportion of labeled data is small, inter-slice augmentation does not achieve segmentation performance comparable to that of fully supervised learning models. In contrast, in Settings 1 and 2, where a larger proportion of data is labeled, the augmentation yields better results. The primary reason for

420this discrepancy is that in Settings 3 and 4, the inter-slice generator is provided with adjacent slices that have a larger gap between them, increasing the distinctiveness of the slices. This makes it more challenging for the generator to accurately predict the intermediate state. Additionally, as more slices were skipped for annotation in these settings, segmentation performance declined sharply, making it harder for the inter-slice augmentation to compensate for this performance loss. This suggests a trade-off between reducing dataset annotation and maintaining the segmentation model's generalization capability.

During model training, we set the FID threshold to 200, with training automatically stopping once the FID drops below this value. This decision stems from pretraining experiments, which identified an FID of 200 as the optimal threshold for boosting segmentation model performance. Setting the FID threshold too low risks mode collapse, where the generator produces a limited variety of images, essentially repeating the same output. For example, when given left- and right-side input images, the model may "cheat" by generating an output identical to the left-side image instead of producing the correct inter-slice image for the middle. This compromises the model's ability to generalize and generate diverse and realistic images. Conversely, if the FID threshold is too high, the generated images may deviate far away from the real ones, reducing their utility. The results in Table 4 validated our choice of the FID threshold, as the segmentation model enhanced with the inter-slice augmentation method outperformed the fully supervised model. Therefore, conducting preliminary experiments to determine the optimal FID threshold is essential before applying this framework.

This study has limitations. First, the dataset used in this study is relatively small, primarily due to the ongoing nature of the cLBP study, with participant enrollment still in progress. Despite the limited sample size, our experiments demonstrated that the dataset is sufficient to develop a preliminary AI model for future analysis. While a larger dataset would likely enhance the model's robustness and generalizability, the current data has provided valuable insights and serves as a solid foundation for future work. Second, the segmentation of the dermis and superficial fat layers is not as accurate as that of other layers. Specifically, the Dice coefficients for the dermis and superficial fat layers are 62.28% and 60.42% respectively (Table 4). This could be attributed to the thinness and close proximity of these two layers, making it challenging to outline their boundaries. Additionally, the implementation of a deblurring model led to a slight, non-significant drop in performance compared to training on a partially annotated dataset. This could be attributed to the model's sensitivity to the diversity of the training dataset. Sharpening the edges of the generated images may reduce the variability of the dataset compared to non-deblurred images. However, for downstream radiomic feature computations, it may be desirable to have images with relatively sharp boundaries to reflect real anatomy. Nevertheless, this deblurring model can be easily removed if it does not improve radiomic feature analysis. Incorporating the deblurring model gives us the option to enhance image clarity when needed.

## 5. Conclusion

We developed and validated InterSliceBoost, a novel deep-learning approach for segmenting tissue layers in 3D B-mode ultrasound images, specifically addressing the challenges associated with manual annotation. By leveraging a partially annotated dataset and generating inter-slice images, our method reduces the burden of manual labeling without compromising segmentation accuracy. The model demonstrated high performance, achieving superior segmentation results with only 33% of the data annotated, outperforming conventional methods trained on fully labeled datasets. This approach not only streamlines the segmentation process but also enables a more comprehensive, layer-by-layer analysis of tissues, providing valuable insights into the structural abnormalities associated with cLBP. As 3D ultrasound becomes a more prevalent imaging modality for studying cLBP biomarkers, InterSliceBoost offers a practical solution for efficiently analyzing large volumes of data while maintaining high-quality segmentation. Future work will focus on expanding the dataset and further improving model generalization to enhance its clinical applicability.